\documentclass[preprint,12pt]{elsarticle}
\usepackage{amsmath,amssymb}
\usepackage[utf8]{inputenc}
\usepackage{amsfonts}
\usepackage{graphicx,xcolor}
\usepackage{bm}
\usepackage{xcolor}
\usepackage{lineno}
\graphicspath{{Figures/}}
\begin{document}

\makeatletter
\def\ps@pprintTitle{%
   \let\@oddhead\@empty
   \let\@evenhead\@empty
   \let\@oddfoot\@empty
   \let\@evenfoot\@oddfoot
}
\makeatother

\begin{frontmatter}

\title{Time-Resolved Vibrational Spectroscopy to Measure Lifetime of the E$^3\Pi_g(v=3)$ state of Molecular Iodine}

\author[First]{Sanjib Thapa}
\author[First]{Lok Pant}
\author[First]{Briana Vamosi\fnref{label1}}
\author[First]{Michael Saaranen}
\author[First]{S. Burcin Bayram\corref{cor1}}
\ead{bayramsb@miamioh.edu}
\fntext[label1]{College of Medicine, University of Cincinnati, USA.}

\cortext[cor1]{Corresponding author}

\address[First]{Department of Physics, Miami University, 500 E. Spring Street, Oxford, OH, USA.}

\begin{abstract}
The lifetime of the $E^3\Pi_g(v=3)$ state of molecular iodine was measured in the gas phase at room temperature. The $E^3\Pi_g$ state was selectively populated by two sequential nanosecond pulse laser excitation. Resolved molecular fluorescence for the $B^3\Pi_u^+\leftarrow E^3\Pi_g$  was analyzed and the lifetime of the $E(v=3)$ state, measured using a time-correlated single photon counting technique, is to be $\tau=21 (2)$ ns.
\end{abstract}


\end{frontmatter}

\linenumbers

\section{\label{sec:level1}Introduction}
\nolinenumbers
\noindent For many decades molecular spectroscopy has provided wide range of opportunities to study the structure and properties of matter in quantum chemistry and fundamental molecular physics. The diatomic iodine spectrum was one of the first spectra to be analyzed due to its simplicity for demonstrating the basic characteristics of diatomic spectra. Studies on molecular iodine have great interests for its various applications due to its spectroscopic convenience and accurately known spectroscopic constants~\cite{Mathieson56,Herzberg50,Herzberg79}. For example, electronically excited molecular iodine has been used many times as a model system for vibrational energy transfer studies~\cite{steinfeld1965,brown1964,steinfeld1970,arnot1958}. The investigation of the properties of iodine is uniquely suited for laser-induced fluorescence (LIF) measurements. Thus, the spectroscopy of molecular iodine is highly important for studying, e.g., radiative lifetime, hyperfine structure for optical frequency metrology and optical communications, time-resolved Raman scattering, optical pumping, and frequency locking of laser standards~\cite{Schmitt97,Williams74,Paisner74,kireev2004,Yan2012,Kobayashi2016,Hrabina2017}. The bridge between the short-range and long-range regions of the interatomic potentials play important role for precision measurements of the oscillatory behavior of wave functions and radiative lifetimes. For instance, depending on the transition between the two electronic orbital symmetries of the molecules the decay rate can be enhanced (superradiant molecule) or destroyed (subradiant molecule) under specific experimental condition. The concept of superradiant has been first introduced by Dicke~\cite{Dicke54} and the measurements of lifetimes, linewidth or a burst of emitted radiation with ringing preceded by a long delay time reveal the nature of the coupling of the internal symmetries of the diatomic molecules~\cite{Macgillivray81}.  \\

\noindent In spite of well over than a century of scientific research, diatomic molecular spectroscopy remains a dynamic area of experimental study.  For instance, astrophysical studies of stellar atmospheres~\cite{Mihalas78} have led naturally to efforts to study spectroscopically exoplanetary atmospheres~\cite{Madhusudhan16,Burrows14}. Development of new techniques, coupled with more established ones, have led to new areas of study associated with cold and ultracold molecules~\cite{McCarron18,Cho11,Zelevinsky06}. In addition to experiments determining spectral line positions, as is frequently done, time dependent and steady state studies measure the coupling between external perturbers such as photons or heavy particle collision partners.  For instance, excitation of excited rovibrational levels with a short pulse laser permits determination of the radiative lifetime of the excited levels.  From these, molecular structural properties such as the internuclear separation dependence of the transition dipole moment may be extracted.  However, measurements of the radiative lifetime or collisional mixing rates for atoms or diatomic molecules seldom can be measured to better than 10\%. This is unfortunate, because such quantities can put stringent limits on the quality of theoretical predictions for the calculations for small molecules.  In this capacity they serve as benchmarks for the calculations for small molecules.   We point out that there are some atomic physics calculations~\cite{Volz96,Jones06} that have quoted uncertainties less than 1\%. \\ 

\noindent Radiative lifetime of an excited state in molecules is  important to understand the molecular spectra and chemical bonds since it can be used as a sensitive probe of perturbations, predissociations, and tunneling effects. Lifetimes are related to the oscillator strengths and transition dipole moments of the  optical transitions and thus are specifically important in molecular astronomy. There have been small number of lifetime measurements in molecules compared to the lifetime measurements in atoms. There are still lack of excited state lifetime measurements in some important ion-molecules, alkali hydride molecules and alkali dimer ion-pair states~\cite{Brzozowski74,Brzozowski76,Brzozowski78,Sanli15,Anunciado16}.  In this study we present for the first time lifetime measurement of the vibrational level of the $E^{3}\Pi_g(v$=$3)$ molecular state of I$_2$ by means of two-step double-resonance pulse excitation spectroscopy. The ground  $X^{1}\Sigma_g^+$ state (rotational constant $B_e$=0.037 cm$^{-1}$ and internuclear distance $r_e$=$2.66~ {\AA}$) and the final $E^{3}\Pi_g$ state (internuclear distance $r_e\approx 3.65~ {\AA}$) have spin forbidden and have the same parity and thus direct excitation is forbidden. The $E^{3}\Pi_g$  state was selectively populated through the intermediate $B^{3}\Pi_u$ state (rotational constant $B_e$=0.029 cm$^{-1}$ and internuclear distance $r_e$=3.024 {\AA})~\cite{Wieland72}. Unfortunately, there are no high vibrational lifetime data available to compare previous experimental study, relevant to our work, by Roussea~\cite{rousseau1975lifetime} who has made lifetime measurement of the $E^{3}\Pi_g(v$=$46)$ state using a two-photon absorption technique. The main purpose of the $E^{3}\Pi_g(v$=$3)$ state lifetime measurement is to compare the previous lifetime measurement to test whether the radiative lifetime of the high vibrational level is similar to that of the low vibrational level, and whether the radiative lifetime is significantly affected by the ion-pair potential. This is specifically important because the $E$-state correlates diabatically with the ground state $I^-(^1S_o)$+$I^+(^3P_2)$ of the separated ions and diabatic dissociation limit corresponds to about 72000 cm$^{-1}$~\cite{Brand82}. Recently, calculated radiative lifetimes for the alkali dimer ion pair states have revealed strong variations as a function of internuclear distance because of the effect of the ion pair potentials~\cite{Sanli15,Michael18,Huwel19}.

\section{Experimental Methods}
\noindent The excitation scheme and partial energy level diagram of iodine molecule~\cite{Mathieson56,Akopyan06,Radzig85} are shown in Fig. 1. The schematic diagram of the experimental setup is shown in Fig 2. We used a pulsed Nd:YAG laser which operates simultaneously at the second harmonics (532 nm) and the third harmonics (355 nm) with a pulse repetition rate of 20 Hz and pulse duration of about 6 ns. The second harmonics of the YAG laser beam (Laser 1), has a bandwidth at FWHM$\approx$ 1~cm$^{-1}$, excites the  $X^{1}\Sigma_g^+(v''=0,J''\approx 53)$$\rightarrow$$B^{3}\Pi_u(v'=32,J'\approx 52)$ transition. The spin-orbit interaction increases as Z$^3$ and thus is much stronger for heavier atoms giving rise to singlet triplet character mixing via spin-orbit interaction. 
The energy of the lowest vibrational level of the triplet state is lower than that of the singlet state. Thus, the vibrational coupling between the ground and excited state occurs and probability of singlet-triplet transition is enhanced. The 355 nm of the YAG, bandwidth at FWHM$\approx$ 0.2~cm$^{-1}$, is used to pump a home-built grazing-incidence Littman-Metcalf cavity oscillator to produce a tunable blue dye laser (Laser 2) at 438.29 nm for the $B^{3}\Pi_u(v'=32,J'\approx 52)$$\rightarrow$$E^{3}\Pi_g(v=3,J\approx 53)$ transition. This dye laser is equipped with a dye circulating system to maintain the lasing and an average power of about 1.3 mW. A cylindrical quartz cell (diameter 25 mm and length 100 mm) contains molecular iodine with background pressure of 2x10$^{-6}$ torr at room temperature. Collimated counter-propagating two pulsed laser beams were directed into the iodine vapor cell in order to minimize the residual Doppler broadening, proportional to the difference in frequency between the two lasers. Both lasers were spatially and temporarily overlapped at the interaction region of the cell. Molecular fluorescence was collected at right angle to the propagation direction of the lasers with a fiber, attached to a wide-angle lens, to efficiently collect the fluorescence which then falls onto the entrance slit of the imaging spectrometer, Horiba iHR320. The entrance and side-exit slits of the spectrometer are set to 50 $\mu$m. The spectrometer eliminates the unwanted signal as it acts as a narrowband filter and prevents stray light.  The spectrometer has two exit ports and each port can be selected with a motorized swing-away mirror. The CCD (Horiba Synergy), has 14 $\mu$m pixel size, for analyzing the molecular fluorescence spectrum and a PMT at the side-exit port for measuring the time-correlated single photon counting  via multichannel scaler (MCS).  

\begin{figure} 
\centering
\includegraphics[scale=0.3]{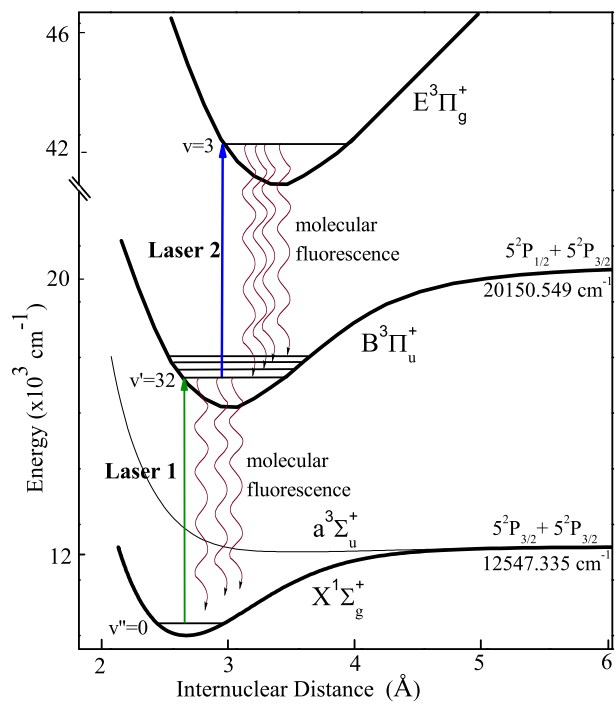}
\caption{\label{fig:epsart} (Color online) Partial potential energy curves of molecular iodine and double-resonance excitation scheme are shown. Dissociation limit of the $E$ state is near 72000 cm$^{-1}$.}
\end{figure}

\begin{figure}[tbh!] 
\begin{center}
{$\scalebox{1.20}{\includegraphics*{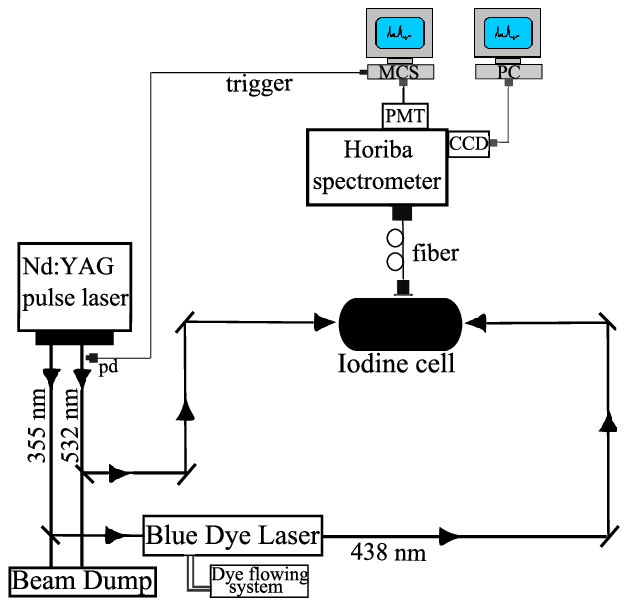}}$ }
\caption{(Color online) A schematic view of the experimental setup. The tunable blue dye laser is pumped by the third harmonics of the pulse Nd:YAG laser which operates simultaneously at 532 nm and 335 nm. The CCD is cooled to -40$^\circ$C, with 14 $\mu$m pixel size. MCS is a multi-channel scaler for time-correlated photon counting, PMT refers to a photomultiplier tube, and pd a photodiode which triggers the MCS. Blue dye laser is a home-built Littrow cavity oscillator and its power is maintained by a dye flowing system.}
\end{center}
\par
\end{figure}

\section{Measurements and Results}

\noindent The laser induced fluorescence for the $X^1\Sigma_g^+(v''=0) \leftarrow B^3\Pi_u(v'=32)$  emission spectrum by laser 1 excitation is shown in Fig. 3. The analysis of this spectrum confirms the selective population in the $B^{3}\Pi_u$($v'$=32) state with intensity peaks match with the Franck-Condon factors~\cite{Tellinghuisen78,Bayram2015,Tellinghuisen2007}. The Franck-Condon factors are calculated using LEVEL Fortran code~\cite{LeRoy07,LeRoy16} which solves the radial Schr$\ddot{o}$dinger equation for bound and quasibound levels numerically, and the output provides Franck-Condon factors, and the Einstein A-coefficients, wave functions, energies of the all possible rovibronic transitions of interest. The $E^3\Pi_g(v=3)$ state is excited by the blue laser (Laser 2). The molecular fluorescence spectrum from the $B^3\Pi_u(v'=32) \leftarrow E^3\Pi_g(v=3)$ transition is analyzed, and Franck-Condon factors were compared to confirm the $E^3\Pi_g(v=3)$ state excitation, shown in Fig.~4. \\

\begin{figure}[tbh!] 
\includegraphics[scale=0.45]{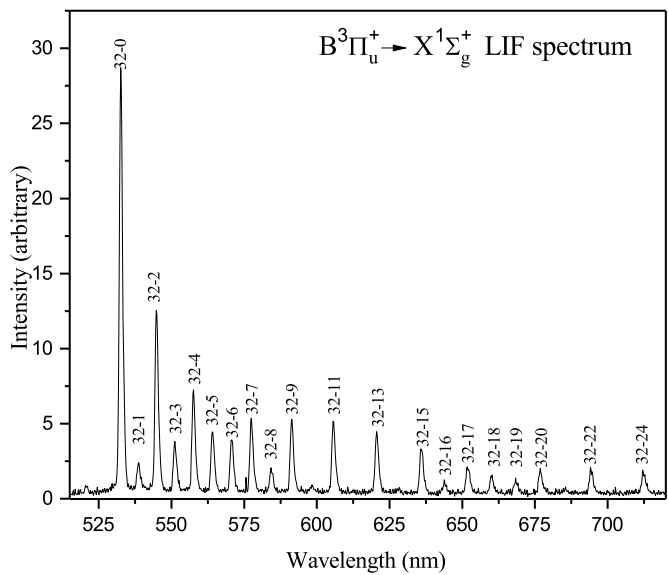}\centering
\caption{Observed laser induced fluorescence (LIF) spectrum for the $X^1\Sigma_g^+(v''=0-24) \leftarrow B^3\Pi_u^+(v'=32)$ emission.}
\end{figure}

\noindent One of the peaks from the spectrum was directed to the PMT and the signal was detected with the multichannel scaler (MCS) to measure the lifetime of the $E^3\Pi_g(v=3)$ state.  The MCS is synchronized with the trigger from the laser pulses and measures the time between the trigger pulse and an associated PMT pulse which results from a single photon. During the scan of 20 000 records were accumulated and displayed on the screen in real time. The result displays a histogram of of fluorescence decay as a function of time. The I$_2$ time decay data was acquired from the MCS for a time-correlated single-photon counting. \\

\begin{figure}[tbh!] 
\centering
\includegraphics[scale=0.3]{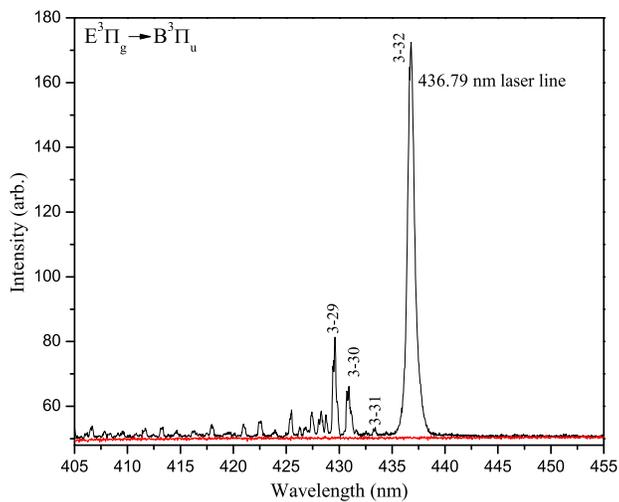}
\caption{(Color online) The laser induced molecular fluorescence for the blue wing of the  $B^{3}\Pi_u^+(v'=32) \leftarrow E^{3}\Pi_g^+(v=3)$ emission is shown. Red line shows the background when laser 1 is blocked.}
\end{figure}

\noindent The exponential fitting of the time decay curve from the MCS data provided the lifetime of the $E^{3}\Pi_g$($v$=3) state. A typical decay curve is shown in Fig.~5. The rotational population distribution of the $X^{1}\Sigma_g^+$ state at room temperature has a maximum at about $J$=52.
The molecular fluorescence from the double-resonance excitation was observed for the $X^1\Sigma_g^+(v''=0,J'' \approx53) \rightarrow B^3\Pi_u(v'=32,J' \approx52) \rightarrow E^3\Pi_g(v=3,J \approx 53)$ transitions. The spectral peaks were identified by comparing them with the output data of the LEVEL Fortran program.  The inverse of the Einstein A-coefficients sum for all possible allowed electric dipole rovibronic transitions, calculated from the output of the LEVEL program, yields the radiative lifetime of the $E$ ion-pair state of I$_2$. We report the lifetime measurement from the (3,29) peak using the time-correlated photon counting method. The Table I lists the lifetimes for the $E^3\Pi_g$ state of I$_2$. \\

\noindent The previously measured lifetimes of the $E(v=0)$ state was 10 (3) ns by Ref.~\cite{King82}, 25.3 (1.4) ns at 100 mTorr and 26.6 (1.4) ns at 50 mTorr by Ref.~\cite{Jewsbury91} who reported no significant pressure difference in the measurements, 28(1) ns by Ref.~\cite{Chevaleyre82} and 25.5 by Ref.\cite{Perrot87}. It is possible for the molecules in the intermediate $B$ state to be excited to a higher rovibrational levels by collisions. The shorter lifetimes maybe related to the excited state I$_2$ collisions with the ground state iodine molecule or due to  weak perturbations in the intermediate $B$ state. 
\begin{figure}[tbh!]  
\includegraphics[scale=0.40]{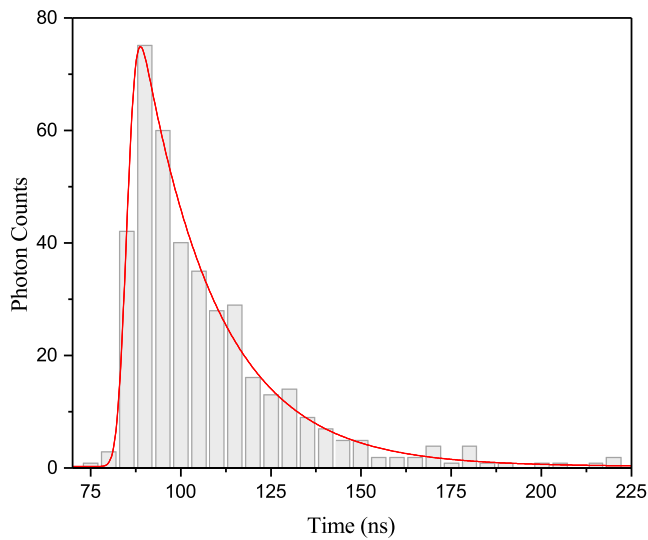}
\centering
\caption{(Color online) Time decay curve of I$_2$ fluorescence for the $E-B$ transition. The photon counting scan illustrating a histogram consisting of a range of time bins from the MCS. The red line is a convolution fit of a Gaussian with an exponential curve function to the data.}
\end{figure}
\begin{table}[h]
\centering
\begin{tabular}{ccccc}
\hline
\textbf{State} & \textbf{Method} & \textbf{Lifetime (ns)} & \textbf{Reference}\\
\hline
$E(3,53)$ & Nd:YAG 5~ns pulse & 21 $\pm2$ & This work\\
$E(46)$   & Kr$^{+}$ 100 ns pulse & 27 $\pm$2 & \cite{rousseau1975lifetime}\\
$E(0)$ & N$_2$ laser  & 10 $\pm$ 3 &\cite{King82}\\
$E(0)$ & XeCl laser  & 25.3 $\pm$ 1.4 &\cite{Jewsbury91}\\
$E(0)$ & XeCl laser  & 28 $\pm$ 1 &\cite{Chevaleyre82}\\
$E(0)$ & Ar$^+$-N$_2$ laser  & 25.5  &\cite{Perrot87}\\
\hline
\end{tabular}
\caption{Tabulated lifetimes measured for the $E^3\Pi_g^+$ electronic state of I$_2$ at various vibrational quantum number at various pressures.}
\end{table}
The presence of impurities in the cell may cause collisional quenching, a relaxation process of excited iodine molecules. This process tends to reduce the lifetime of the excited state, causes non-radiative transitions of iodine. In addition, laser power may affect the lifetime measurements. We have tested the laser power dependency on the lifetime measurements and the data was collected for the linear regime, power-dependency free. In addition, collisions among iodine atoms and molecules and background gas collisions affect the lifetime measurements.  This can happen via a number of quenching and energy transfer processes including rotational or vibrational energy transfer.  Collisions transferring atoms to lower energy electronic states can also be important.  To estimate the role of such processes in our experiment, we require at least an approximate value for the background gas pressure in the sample cell.  With such an estimate, and an approximate value for the collision process cross section, the collision rate may be found from a Stern-Volmer type equation.  When that rate is much smaller than the radiative decay rate, collisional processes may be safely ignored. However, this is a difficulty for sealed cells such as we use in experiments on I$_2$.  To judge collisional processes we need the gas pressure in the cell.  We estimated the iodine pressure from the vapor pressure curves for this molecule to be approximately 0.1 Torr at room temperature.  This pressure can also be varied by cooling the sidearm on the iodine cell.  Estimating the effect of outgassing of the Pyrex cell requires an approximate value for the outgassing rate and the initial background gas pressure in the cell.  The manufacturer~\cite{iodine} specification for the initial background pressure is 2 x 10$^{-6}$ Torr.  A characteristic outgassing rate for Pyrex is 10$^{-8}$ Torr-liters/s/cm$^2$. As the outgassing pressure depends on time (it is cumulative), we take a characteristic time scale for the measurements of 10$^6$ s. As a result, we estimated the background pressure to be dominated by the iodine atom vapor pressure, the influence of which has been tested experimentally by varying the cell temperature, and which is negligible at room temperature.


\section{Conclusion}
\noindent The experimental study of the lifetimes is important to interpret the excited state molecular structure. In this work, we present new experimental data on the $E(v=3,J=52,53)$ state lifetime of the diatomic iodine. Our experimental lifetime value is slightly lower than the measured lifetime of the higher vibrational $E(v=49)$ state~\cite{rousseau1975lifetime}. However, within the errors,  we conclude that the lifetime is not significantly affected by the ion pair potential. Lifetimes or decay rates as well as Doppler free linewidths may reveal the nature of the states involved by the scale of these quantities in relation to the long range interatomic interactions. For instance, in strontium ultracold homonuclear molecules, the short lifetime of the $1_u$ state implies that the state is equivalent to a two body superradiant state~\cite{McGuyer15}.  On the other hand, measurements of a long sequence of lifetimes as a function of ($v,J$) can reveal scaling laws in these quantities. Recently published articles reported a strong dependence of molecular lifetime on rovibrational level's wavefunctions~\cite{Michael18,Huwel19}. This work will be followed by making series of lifetime measurements as a function of vibrational or rotational quantum numbers in diatomic molecules.

\section*{Acknowledgments}
\noindent Financial support from the National Science Foundation (Grant No. NSF-PHY-1607601) is gratefully acknowledged.


\end{document}